\documentclass[final,1p,times]{elsarticle}

\usepackage{amssymb}
\usepackage{amsmath}
\usepackage{color}
\usepackage{graphicx}
\usepackage{array}

\begin{document}

\begin{frontmatter}

\title{Promotion and resignation in employee networks}

\author[inst1]{Jia Yuan}
\author[inst2,inst3]{Qian-Ming Zhang \corref{cor1}}
\author[inst2,inst4]{Jian Gao}
\author[inst4]{Linyan Zhang}
\author[inst6]{Xue-Song Wan}
\author[inst6]{Xiao-Jun Yu}
\author[inst2,inst5]{Tao Zhou}
\cortext[cor1]{Correspondence author: qmzhangpa@gmail.com}

\address[inst1]{School of Management and Economics, University of Electronic Science and Technology of China, Chengdu 611731, People's Republic of China}
\address[inst2]{CompleX Lab, Web Sciences Center, School of Computer Science and Engineering, University of Electronic Science and Technology of China, Chengdu 611731, People's Republic of China}
\address[inst3]{Center for Polymer Studies, Department of Physics, Boston University, Boston 02215, United States of America}
\address[inst4]{Hire Big Data (Chengdu) LTD., Chengdu, People's Republic of China}
\address[inst5]{Big Data Research Center, University of Electronic Science and Technology of China, Chengdu 611731, People's Republic of China}
\address[inst6]{Beijing Strong Union Technology Co. Ltd., Beijing, People's Republic of China}

\begin{abstract}
  Enterprises have put more and more emphasis on data analysis so as to obtain effective management advices. Managers and researchers are trying to dig out the major factors that lead to employees' promotion and resignation. Most previous analyses are based on questionnaire survey, which usually consists of a small fraction of samples and contains biases caused by psychological defense. In this paper, we successfully collect a data set consisting of all the employees' work-related interactions (action network, AN for short) and online social connections (social network, SN for short) of a company, which inspires us to reveal the correlations between structural features and employees' career development, namely promotion and resignation. Through statistical analysis, we show that the structural features of both AN and SN are correlated and predictive to employees' promotion and resignation, and the AN has higher correlation and predictability. More specifically, the in-degree in AN is the most relevant indicator for promotion, while the $k$-shell index in AN and in-degree in SN are both very predictive to resignation. Our results provide a novel and actionable understanding of enterprise management and suggest that to enhance the interplays among employees, no matter work-related or social interplays, can largely improve the loyalty of employees.
\end{abstract}


\end{frontmatter}

\section{Introduction}
Employees are the core and soul of enterprises, and thus human resources departments are always trying to get comprehensive understanding of employees and provide to managers valuable and effective suggestions. This issue has attracted much attention from interdisciplinary research domains. Some factors related to the employees' performance have been revealed, such as centrality \cite{Ahuja2003,Mehra2001}, self-monitoring orientation \cite{Ahuja2003}, sex \cite{Bu2005}, and communication patterns \cite{Pentland2012}. Besides the performance, predicting and controlling resignation in advance are also valuable, since employees' resignation may result in great losses for enterprises. Freely and Barnett \cite{Feeley1997} showed that people being highly connected and in more central positions of the employee networks are less likely to resign. Ten years later, this group put forward that employees who reported a greater number of out-degree links to friends are less likely to resign \cite{Feeley2008}. Similar indicators include degree, betweenness and closeness \cite{Mullen1991,Mossholder2005}.

Traditional studies are based on the data sets gathered from questionnaire survey that is probably subjective due to the psychological defense. Fortunately, recent IT development brings us more reliable information, such as email data \cite{Guimera2003}, communication records \cite{Eagle2006}, online social relations \cite{Lewis2012}, contact networks built by wearable electronic sensors \cite{Pentland2012}, and so on. In this paper, we collect anonymous employees' work-related interactions and social connections from a social network platform developed and used by a Chinese company consisting of more than a hundred employees, named Beijing Strong Union Technology Co. Ltd. (Strong Union for short). Accordingly, we can build two directed networks, action network (AN) and social network (SN), where nodes represent employees and links indicate the work-related interactions and social connections, respectively. A few studies have already shown that the employees being more central in the networks or with more connections to others are less likely to resign \cite{Feeley1997, Feeley2008, Mullen1991, Mossholder2005, McPherson1992, Moynihan2008}, while managers whose local networks are rich in structural holes could be promoted faster \cite{Burt1992}. Here, we take more structural features into consideration to reveal the correlations between topology and career changes. In addition, we would like to show which network is more related to promotion or resignation.

The two networks characterize relationships among employees from different aspects. AN reflects work-related interactions such as downloading working files and working blogs within working groups, while SN reflects social connections among all employees of the company, through which employees can share their family and entertainment activities (usually not related to works) with others. Via statistical analysis, we find that the structural features of both AN and SN are correlated and predictive to employees' promotion and resignation, and the AN has higher correlation and predictability. More specifically, the in-degree in AN is the most relevant indicator for promotion, while the $k$-shell index in AN and in-degree in SN are both very predictive to resignation.

This paper is organized as follows. Section 2 presents the description and basic statistics of the data set. The significance of selected structural features and logistic regression are respectively showed in Section 3 and Section 4. At last we give our conclusions and discussion in Section 5.

\section{Data Description}
The data set is gathered from a social network platform which involves all the 104 employees in Strong Union until the end of 2013. It consists of two types of information, work-related actions and online social connections. The work-related actions of employees include reposting and commenting working blogs, assigning and reporting working tasks, and downloading working files. The online social connections, which are similar to the follower-followee relationships in \emph{www.twitter.com}, are created regardless of working groups, namely everybody can build relationships with others even they don't belong to any common working groups. Accordingly, we can build two directed networks, action network (AN) and social network (SN), where nodes represent employees and links indicate the work-related interactions and social connections, respectively. More specifically, a link from $u$ to $v$ in AN means the employee $u$ has reposted, commented or downloaded $v$'s working files, and a link from $u$ to $v$ in SN means $u$ has followed $v$. To the date of the data collection, 25 employees have resigned and 12 employees were promoted.

Table \ref{tableDataDes} summarizes the basic topological features of the two networks. Compared with AN, SN is of higher density and average degree. The two networks are both of high clustering coefficients, quite small average shortest path lengths and very small modularity, since they are so dense that all the nodes are grouped together. Different from many known social networks \cite{Newman2002}, the two networks are both highly disassortative. It is because many work-related interactions and social connections happen between leaders and ordinary employees: the former are usually of large degrees while the latter are usually less connected.

\begin{table*}[htbp]
\begin{center}
  \caption{The basic topological features of the two networks. $N$ is the total number of nodes. $D$ denotes the density of directed networks calculated by $|E|/N(N-1)$, where $|E|$ is the number of directed links. Other features are calculated by transferring the directed networks into undirected ones.  $\langle k\rangle$ is the average degree,  $\langle d\rangle$ is the average shortest path length,  $r$ is the assortative coefficient \cite{Newman2002}, and  $C$ is the clustering coefficient \cite{Watts1998}.  $H$ is the degree heterogeneity quantified by a variant of Gini index \cite{Hu2008}: The higher $H$ is, the more heterogeneous the degree distribution is.  $Q$ is the modularity of a network calculated by the Louvain method \cite{Blondel2008}.}
  {\begin{tabular}{c|cccccccc}
  \hline\hline
   Networks & $N$ & $D$ & $\langle k\rangle$ & $\langle d\rangle$ & $r$ & $C$ & $H$ & $Q$ \\
  \hline
   AN & 97	&0.26	&35.73	&1.64	&-0.27	&0.76	&0.35	&0.09 \\
   SN & 104	&0.29	&47.04	&1.55	&-0.41	&0.81	&0.32	&0.08 \\
\hline\hline
\end{tabular}\label{tableDataDes}}
\end{center}
\end{table*}

\section{Significance of Structural Features}
Previous studies suggested that some structural features of employees are effective to evaluate the resignation \cite{Feeley1997, Feeley2008, Mullen1991, Mossholder2005, McPherson1992, Moynihan2008}. Here we select three well-known indicators in-degree ($k_i$), out-degree ($k_o$) and $k$-shell index ($k_s$). Given a network $G$, $k_i$ is the number of links pointing to a node and $k_o$ is the number of a node's outgoing links. $k_s$ of a node is the largest number such that this node belongs to the $k_s$-core, which is the maximum subgraph of $G$ in which all nodes have degree no less than $k_s$ \cite{Seidman1983}. In the calculation of $k_s$, the networks are treated as undirected ones. This index defines the employees' positions in the network: the employees with higher $k_s$ are more central in the network, and vice versa. The $k$-shell index and its variants have been used to identify the influential spreaders in social networks \cite{Kitsak2010, Liu2013, Zeng2013, Chen2012, Liu2014}.

Because the career changes can be treated as binary variables, i.e. promotion vs. non-promotion and resignation vs. non-resignation, we adopt AUC (area under the receiver operating characteristic curve \cite{Hanely1982}) value to measure the correlation between structural features and career changes. This metric can be interpreted as the probability that a randomly chosen promoted (non-resigned) employee has a higher value of  $k_i$, $k_o$, or $k_s$ than a randomly chosen non-promoted (resigned) employee (see the definition and detailed explanation in \cite{Lv2012}). In the implementation, we compare all the pairs of employees with opposite states (promotion vs. non-promotion and resignation vs. non-resignation) to calculate AUC. Suppose that among $m$ times of comparisons in total, there are $m'$ times the promoted (non-resigned) employee has higher value than the non-promoted (resigned) employee, and $m''$ times they are equal, the AUC value can be calculated by
\begin{equation}\label{AUCvalue}
  \mathrm{AUC}=\frac{m'+0.5m''}{m}.      \nonumber
\end{equation}
If the topological features are independent to the career changes, the AUC value should be about 0.5. Therefore, the degree to which the value exceeds 0.5 indicates the strength of correlation.

\begin{figure*}[ht]
\centering
  \includegraphics[width=11cm]{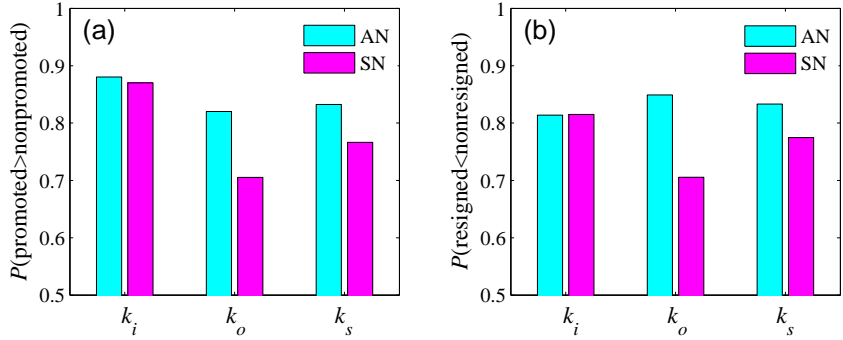}
  \caption{(Color online) AUC values for promotion and resignation. Subfigure (a) represents the probability that a randomly chosen promoted employee has a higher value of $k_i$, $k_o$ or $k_s$ than a randomly chosen non-promoted employee. Subfigure (b) represents the probability that a randomly chosen resigned employee has a lower value of $k_i$, $k_o$ or $k_s$ than a randomly chosen non-resigned employee.}\label{fig-AUC}
\end{figure*}

As indicated by the high AUC values in Fig. \ref{fig-AUC}, promoted employees have obviously higher values for all the three structural metrics, while resigned employees have much lower values than others. These results strongly imply that the employees who are more central or highly connected are more likely to be promoted and less likely to resign. In addition, there are three secondary results: (i) AN is more related to both promotion and resignation than SN; (ii) in-degree in AN, namely the number of colleagues who have responded to the target employee's working files, is the most critical indicator for promotion; (iii) out-degree in AN, indicating the number of employees the target employee follows, is the most relevant indicator for resignation: The more you follow, the less likely you quit.

\section{Logistic Regression}
Since employees are only with two states in each case (promotion vs. non-promotion and resignation vs. non-resignation), we adopt logistic regression \cite{Bishop2006} to identify the promoted employees and resigned employees. It is a type of probabilistic statistical classification model, which is usually used to predict the binomial outcome of a response variable using one or more predictor features (e.g., the three indicators in this paper). The binary logistic regression model can be represented by a conditional probability
\begin{equation}\label{LogRegModel}
  P(1|\overrightarrow{x})=\frac{1}{1+e^{-(b_0+\sum_{i}^{m}b_ix_i)}},
\end{equation}
where $\overrightarrow{x}=(x_1,...,x_m)$ is the vector of features, such as $k_i$, $k_o$ and $k_s$, and $b_0,b_1,...,b_m$ are the coefficients we need to estimate based on the data. Here we use 1 to indicate promotion and resignation, and 0 to indicate non-promotion and non-resignation.

By assigning $b_0,b_1,...,b_m$ proper values, we can calculate the probability $P(1|\overrightarrow{x})$ through Eq. (\ref{LogRegModel}) for each employee. Then we are able to divide the employees into two groups according to a chosen cutoff probability (which is usually set to 0.5 by default). One group corresponds to promotion (resignation), and the other corresponds non-promotion (non-resignation). We assume a model is better if more employees are identified correctly, namely, it is more capable to identify which employees are promoted (resigned) and which are non-promoted (non-resigned). To measure the accuracy of classification, we firstly introduce a basic metric \emph{precision}, defined as
\begin{equation}
  precision=\frac{n}{N},     \nonumber
\end{equation}
where $N$ is the total number of employees and $n$ is the number of correctly classified employees. Nevertheless, only considering \emph{precision} is not good enough in this case, because the numbers of both promoted employees and resigned employees are very small. Then we can get high precision by simply classifying all the employees into the ¡°non-promoted¡± or ¡°non-resigned¡± class (up to 88.5\% for promotion and 76\% for resignation), but no relevant employees (i.e., promoted employees and resigned employees) can be found out. Fortunately, \emph{recall}, which is known as the sensitivity in binary classification, can compensate that inadequate measurement. The mathematic form is defined as
\begin{equation}
  recall=\frac{n'}{N'},     \nonumber
\end{equation}
where $n'$ is the number of relevant employee who is retrieved, and $N'$ is the number of relevant employees who have resigned or have been promoted.\\

\begin{figure*}[ht]
\centering
  \includegraphics[width=10.5cm]{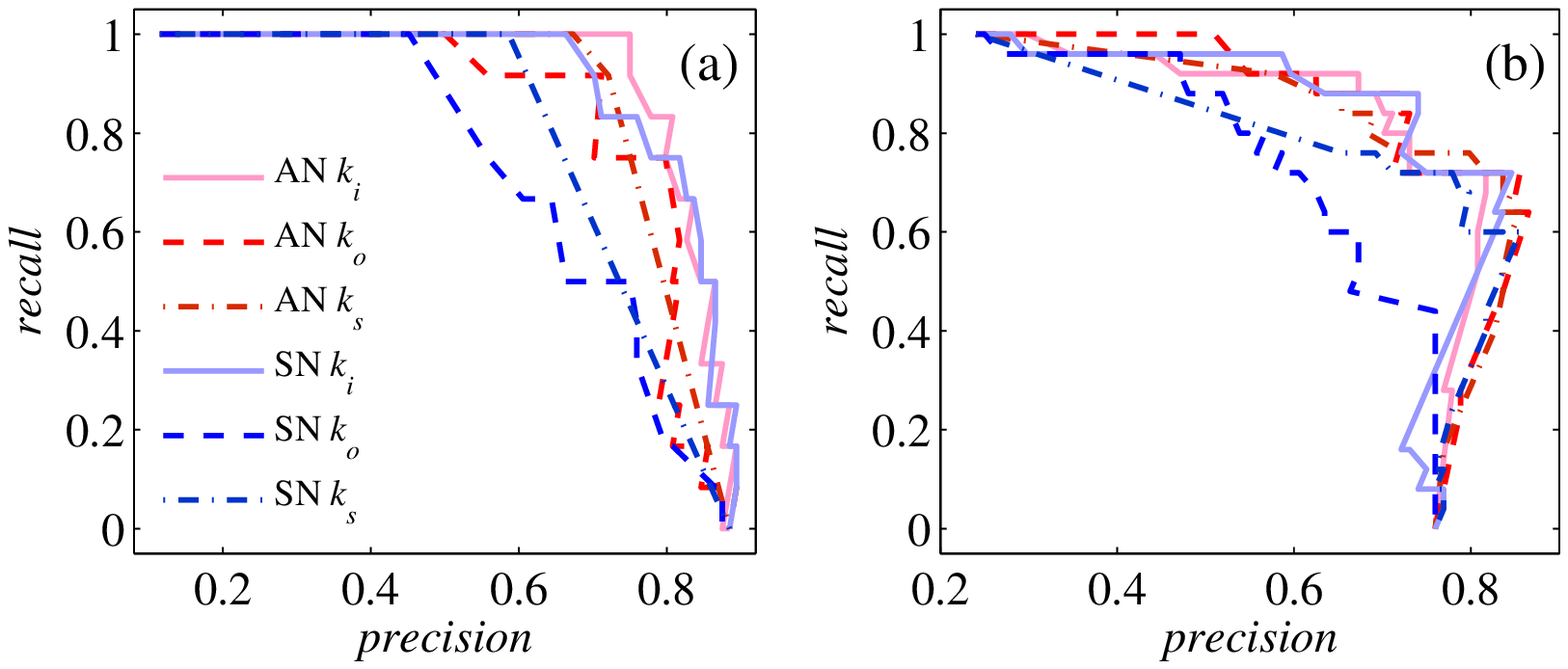}
  \caption{(Color online) \emph{Precision-recall} curve with various cutoff probabilities. Subfigure (a) represents the \emph{precision-recall} curves for promotion, and subfigure (b) represents the curves for resignation. AN represents action network, and SN represents social network. $k_i$, $k_o$ and $k_s$ are the selected variables to classify which employees are promoted or resigned. The significance of chi-square test [28] for the constant $b_0$ and coefficient $b_1$ according to the six models for promotion are (0, 0), (0.002, 0), (0.221, 0.208), (0, 0), (0.058, 0), and (0.987, 0.987) respectively. And those for resignation are (0, 0.755), (0, 0.164), (0, 0.022), (0, 0.021), (0.009, 0.054) and (0, 0.007) respectively. In general, the value below 0.05 shows a good fit.}\label{fig-PreRec}
\end{figure*}

\begin{figure*}[ht]
\centering
  \includegraphics[width=10.5cm]{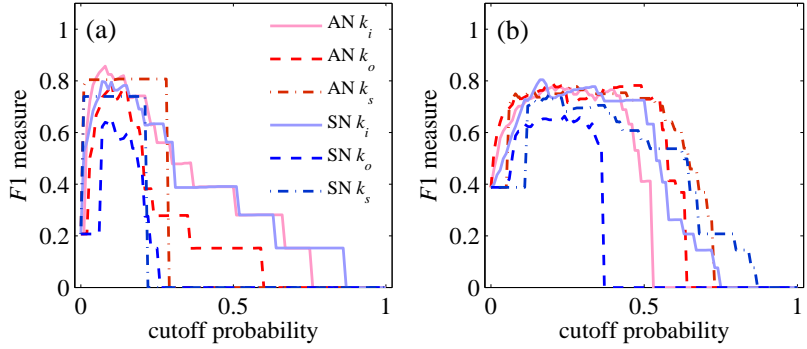}
  \caption{(Color online) $F$1 scores with various cutoff probabilities. Subfigure (a) represents the $F$1 scores for promotion, and subfigure (b) represents the $F$1 scores for resignation. AN means action network, and SN means social network. $k_i$, $k_o$ and $k_s$ are the selected variables to classify which employees are promoted or resigned.}\label{fig-F1}
\end{figure*}

To figure out which feature is most related to promotion/resignation, in each run, we only choose one feature to form the vector $\overrightarrow{x}$. The one with higher \emph{precision} and \emph{recall} is considered to be more related. Need to notice that, we do not put the three features in one model, because these three features are highly correlated with each other. And we do not compare the estimated coefficients for these features, because it will lead us inaccurate conclusions \cite{Mood2010}. Apparently, the values of \emph{precision} and \emph{recall} depend on the cutoff probability. So, if we vary the cutoff probability from 0 to 1, we can get a \emph{precision}-\emph{recall} curve. As shown in Fig. \ref{fig-PreRec}, the most accurate classification can be found according to the right upper curve. More directly and clearly in visualization, we introduce another metric named $F$1 \cite{Rijsbergen1979, Levis1994} to compare the effectiveness by synthesizing \emph{precision} and \emph{recall} in the following way:
\begin{equation}
  F1=\frac{2\times precision\times recall}{precision+recall},     \nonumber
\end{equation}
It can be interpreted as a weighted average of the \emph{precision} and \emph{recall}. Obviously, an $F$1 score reaches its best value at 1 and worst score at 0. The $F$1 scores for individual features under different cut-off probabilities can be found in Fig. \ref{fig-F1}.\\

From the peak values of the curves for promotion, we can see that $k_i$ is the most effective feature, while $k_o$ is the worst. In such small-scale networks, $k_i$ can reflect the employee's direct influence since the leaders are more likely to attract more attention. Following this idea, $k_s$ should be also valid, because the higher $k_s$ an employee has, the more central position he stands. However, $k_s$ does not work so well. It may be because there are too many employees with the highest $k_s$, i.e., 39 out of 97 employees with highest $k_s$ in AN, and 55 out of 104 employees with highest $k_s$ in SN. These employees own quite different values of $k_i$ ranging from 33 to 82 for AN, and from 37 to 68 for SN. So the employees are more distinguishable in terms of $k_i$ than that of $k_s$. Compared with SN, every corresponding feature of AN is better. Further, let us check the significance of these estimated coefficients (see the caption of Fig. \ref{fig-PreRec}). Only $k_i$ in AN, $k_o$ in AN, and $k_i$ in SN can be accepted. So we can conclude that in-degree in the action network is the most effective and acceptable feature for predict promotion.

In the case of resignation, although the maximal $F$1 score belongs to $k_i$ in SN, the three features of AN are all very competitive. These results make sense because employees who get little attention (low $k_i$ in SN and AN), or stands at the periphery of the social network (low $k_s$ in SN) would feel worthless, under-appreciated and lonely. These negative feelings would results in resignation. In a word, employees who own many followers (high $k_i$), stand at important positions (high  $k_s$), or are the hubs of communication (AN  $k_o$) are less likely to resign. When we check the significance of these estimated coefficients, only $k_s$ in AN, $k_s$ in SN and $k_i$ in SN are acceptable. So we conclude that $k_s$ in AN and $k_i$ in SN are better.

Through the above analysis, we can conclude $k_i$ in AN is most related to promotion, while $k_s$ in AN and $k_i$ in SN are most related to resignation.

\section{Conclusions and Discussion}
In this paper, we successfully collect a dataset consisting of all the employees' work-related actions and online social connections from a social network platform developed and used by a company. To reveal the correlation between employees' social/work-related actions and employees' career development, we implement some experiments based on the structural features of those two networks. The correlation analysis shows that the work-related interactions are more correlated to both promotion and resignation, which is also consistent with the classification results through Logistic Regression analysis. We find the employee who can get more attention (higher in-degree) in the work-related network, is more probable to be promoted, while the employees with low values of $k$-shell index in the work-related network and in-degree in the social network have high risk to resign. It makes sense that peripheral employees (with low $k$-shell value) are more likely to leave, because they may feel worthless, under-appreciated and lonely. In turn, we may think the central employees (with high $k$-shell value) will get higher probability to be promoted, but the classification experiments give negative results, which may be resulted from the fact that there are too many employees with the highest $k$-shell value. The results provide a novel and actionable understanding of enterprise management and suggest that to enhance the interplays among employees, no matter work-related or social interplays, can largely improve the loyalty of employee.

In these two employee networks, employees are so highly connected with each other that the networks are not obvious with modules. Maybe that is the nature of small-scale companies, but we can imagine that connections might be very sparse and the corresponding network might be organized with community structure in large-scale companies. Moreover, Hedstrom \cite{Hedstrom1991} had indicated that the organization structures are usually different in different countries. So the current conclusions may be only valid in small-scale enterprises and cannot be directly extended to organizations with different sizes, in different domains or belonging to different cultures. However, our conclusions are reasonable. We suggest human resources to pay more attention to employees' behaviors, e.g., the one who is socially isolated, and the companies also need to appreciate the ordinary employees who are in the central positions, because they are good organizers who can help senior leaders to unite the company.

Furthermore, promotion and resignation are also affected by various reasons. For example, maybe the employees in the central position have already been the senior leaders who are restricted by ``vacancy chains'' \cite{Chase1991}. Some employees leave the company just because of the expiry of the employment contract. In despite of its complexity, we believe that the present work and further quantitative analyses on non-intervention data could provide novel insights in addition to those from questionnaire survey. We hope it could inspire further investigation on related issues to complement the current conclusions.

\section*{Acknowledgments}
We are grateful to Jin Yin for his valuable comments and suggestions. This work is partially supported by the National Nature Science Foundation of China under Grand No.11222543. QMZ acknowledges China Scholarship Council and the support from the Program of Outstanding PhD Candidate in Academic Research by UESTC (No. YBXSZC20131034).


\section*{References}
\begin{thebibliography}{99}
\bibitem{Ahuja2003} M.K. Ahuja, D.F. Galletta, K.M. Carley, Individual centrality and performance in virtual R\&D groups: An empirical study, Manage. Sci. 49 (2003) 21.

\bibitem{Mehra2001} A. Mehra, M. Kilduff, D.J. Brass, The social networks of high and low self-monitors: Implications for workplace performance, Admi. Sci. Quart. 46 (2001) 121.

\bibitem{Bu2005} N. Bu, J.P. Roy, Career success networks in China: sex differences in network composition and social exchange practices, Asia Pac. J. Manage. 22 (2005) 381.

\bibitem{Pentland2012} A.S. Pentland, The new science of building great teams, Harvard Bus. Rev. 90 (2012) 1.

\bibitem{Feeley1997} T.H. Feeley, G.A. Barnett, Predicting employee turnover from communication networks, Hum. Commun. Res. 23 (1997) 370.

\bibitem{Feeley2008} T.H. Feeley, J. Hwang, G.A. Barnett, Predicting employee turnover from friendship networks, J. Appl. Commu. Res. 36 (2008) 56.

\bibitem{Mullen1991} B. Mullen, C. Johnson, E. Salas, Effects of communication networks structure: Components of positional centrality, Soc. Networks 13 (1991) 169.

\bibitem{Mossholder2005} K.W. Mossholder, R.P. Settoon, S.C. Henagan, A relational perspective on turnover: Examining structural, attitudinal, and behavioral predictors, Acad. Manage. J. 48 (2005) 607.

\bibitem{Guimera2003} R. Guimer\`a, L. Danon, A. D\'{i}az-Guilera, F. Giralt, A. Arenas, Self-similar community structure in a network of human interactions, Phys. Rev. E 68 (2003) 065103.

\bibitem{Eagle2006} N. Eagle, A.S. Pentland, Reality mining: sensing complex social systems, Pers. Ubiquit. Comput. 10 (2006) 255.

\bibitem{Lewis2012} K. Lewis, M. Gonzalez, J. Kaufman, Social selection and peer influence in an online social network, Proc. Natil. Acad. Sci. USA 109 (2012) 68.

\bibitem{McPherson1992} J.M. McPherson, M. Popielarz, S. Drobnic, Social networks and organizational dynamics, Am. Sociol. Rev. 57 (1992) 153.

\bibitem{Moynihan2008} D.P. Moynihan, S.K. Pandey, The ties that bind: Social networks, person-organization value fit, and turnover intention, J. Public. Admin. Res. Theory 18 (2008) 205.

\bibitem{Burt1992} R.S. Burt, Structural holes: The structural holes of competition, Cambridge: Harvard University Press, 1992.

\bibitem{Newman2002} M.E.J. Newman, Assortative mixing in networks, Phys. Rev. Lett. 89 (2002) 208701.

\bibitem{Watts1998} D.J. Watts, S. H. Strogatz, Collective dynamics of `small-world' networks, Nature 393 (1998) 440.

\bibitem{Hu2008} H.-B. Hu, X.-F. Wang, Unified index to quantifying heterogeneity of complex networks, Physica A 387 (2008) 3769.

\bibitem{Blondel2008} V.D. Blondel, J.L. Guillaume, R. Lambiotte, E. Lefebvre, Fast unfolding of communities in large networks, J. Stat. Mech. 2008 (2008) P10008.

\bibitem{Seidman1983} S.B. Seidman, Network structure and minimum degree, Soc. Networks 5 (1983) 269.

\bibitem{Kitsak2010} M. Kitsak, L.K. Gallos, S. Havlin, F. Liljeros, L. Muchnik, H. E. Stanley, H. A. Makes, Identification of influential spreaders in complex networks, Nat. Phys. 6 (2010) 888.

\bibitem{Liu2013} J.-G. Liu, Z.-M. Ren, Q. Guo, Ranking the spreading influence in complex networks, Physica A 392 (2013) 4154.

\bibitem{Zeng2013} A. Zeng, C.-J. Zhang, Ranking spreaders by decomposing complex networks, Phys. Lett. A 377 (2013) 1031.

\bibitem{Chen2012} D. Chen, L. L\"u, M.-S. Shang, Y.-C. Zhang, T. Zhou, Identifying influential nodes in complex networks, Physica A 391 (2012) 1777.

\bibitem{Liu2014} Y. Liu, M. Tang, T. Zhou, Y. Do, Core-like groups resulting in invalidation of k-shell decomposition analysis, arXiv: 1409.5187, (2014).

\bibitem{Hanely1982} J.A. Hanely, B.J. McNeil, The meaning and use of the area under a receiver operating characteristic (ROC) curve, Radiology 143 (1982) 29.

\bibitem{Lv2012} L. L\"u, T. Zhou, Link prediction in complex networks: A survey, Physica A 390 (2012) 1150.

\bibitem{Bishop2006} C.M. Bishop, Pattern Recognition and Machine Learning, Vol. 1. New York: Springer, 2006.

\bibitem{Yates1934} F. Yates, Contingency table involving small numbers and the $\chi^2$ test, Supp. J. Roy Stat. Soc. 1 (1934) 217.

\bibitem{Mood2010} C. Mood, Logistic regression: Why we cannot do what we think we can do and what we can do about it, Eur. Sociol. Rev. 26 (2010) 67.

\bibitem{Rijsbergen1979} C.J. van Rijsbergen, Information Retrieval, London: Butterworths, second edition, 1979.

\bibitem{Levis1994} D.D. Lewis, W.A. Gale, A sequential algorithm for training text classifiers, in:  Proceedings of the 17th annual international ACM SIGIR conference on Research and development in information retrieval (SIGIR '94), Springer-Verlag, New York, 1994.

\bibitem{Hedstrom1991} P. Hedstrom, Organizational differentiation and earnings dispersion, Am. J. Sociol. 97 (1991) 96.

\bibitem{Chase1991} I.D. Chase, Vacancy Chains, Annu. Rev. Sociol. 17 (1991) 133.


\end{thebibliography}
\end{document}